\documentstyle[preprint,aps,prl,eqsecnum]{revtex} 
\begin{document} 
\draft 
\preprint{pre3.tex} 
\title{ 
  Ultrasound scattering and the study of vortex correlations in disordered  
  flows 
} 
\author{ 
  {\sc  Denis Boyer \footnote{Present address: Supercomputer Computations
Research Institute, Florida State University, Tallahassee, Florida
32306-4130.} and Fernando Lund} 
} 
\address{  
    {\it Departamento de F\'\i sica, 
    Facultad de Ciencias F\'\i sicas y Matem\'aticas, \\ 
    Universidad de Chile, 
    Casilla 487-3, Santiago, Chile.} 
} 
\date{\today} 
\maketitle 
\begin{abstract} 
In an idealized way, some turbulent flows can be pictured by assemblies of 
many vortices characterized by a set of particle distribution functions. 
Ultrasound provide an useful, nonintrusive, tool to study the  
spatial structure of vorticity in flows. This is analogous to the use of 
elastic neutron scattering to determine liquid structure. 
We express the dispersion relation, 
as well as the scattering cross section, of sound waves  
propagating in a ``liquid'' of  
identical vortices as a function of vortex pair correlation functions.  
In two dimensions, formal analogies with ionic liquids are pointed out. 
 
\end{abstract} 
 
\section{Introduction} 
 
Neutron scattering has been an essential experimental tool 
for the study of liquid structure for the past 30 years \cite{lovesey}.  
The fact that, in a first Born approximation, the scattering cross section 
is directly proportional to the static structure factor 
has been a source both of data and inspiration for the elaboration 
and verification of current theories of liquid structure \cite{scatt}. 
 
At first sight, a turbulent fluid bears no relation to the microscopic 
structure of a liquid. However, a turbulent fluid has vortices, and these 
objects couple to acoustic waves in much the same way as neutrons, 
or electromagnetic waves, couple to matter. The purpose of the present 
paper is to point out a number of theoretical developments that make  
this analogy more precise, particularly with respect to ionic liquids. 
 
It is well known experimentally \cite{belin,cadot,abry} as well as 
numerically 
\cite{siggia,she,vincent,tanaka,jimenez} that coherent vortical structures 
easily appear 
in turbulent flows.  
Three-dimensional turbulent flows of high Reynolds number 
exhibit many intense, long-lived and tube-like vorticity regions 
of fairly well defined thickness and length \cite{jimenez}.  
These filaments have a tendency to organize parallel one to each other in
clusters. 
Although they generally represent a small part of the motion of the fluid, 
the filaments probably contain information on the statistics of the whole 
background. 
In a quite different context, 
numerical studies of two-dimensional decaying turbulence also reveal the  
emergence of coherent vortices with particle-like character. 
These vortices dominate the long-time motion of the  
fluid. Their size and number density are well described by scaling  
laws, and their dynamics are similar to the  
Hamiltonian motion of few point vortices \cite{benzi,carnevale,watanabe}. 
According to these observations, two-dimensional decaying turbulence can be  
approximately  
described with a finite number of degrees of freedom. Following this 
approach,  
statistical mechanical theories of the Euler's equation in a bounded domain 
have been developed for systems of point  
vortices \cite{onsager,novikov,pointin}, and extended afterwards to 
continuous  
distributions of vorticity \cite{rosom,miller}.  
These theories where able to explain the results of two-dimensional  
Hamiltonian simulations \cite{montgomery}, 
showing that point vortices of same circulation sign are not randomly mixed, 
but have a tendency to organize in a nonuniform way within the domain. 
In infinite domains, homogeneous two-dimensional turbulence appears to be 
well  
described by many large coherent vortices that must be spatially correlated. 
Quantitative experimental or numerical studies of these correlations are 
few, 
although two dimensional turbulence has become in recent years the object of
controlled laboratory experiments \cite{soap,tabeling}
 
Ultrasound scattering provides a powerful, nonintrusive tool to  
study vortical structure in flows.  
The characterization of the scattered pressure of an ultrasonic plane wave at  
sufficiently large distances from a vorticity distribution has motivated  
various theoretical studies 
\cite{obukhov,kraichnan,howe,fabricant,lundrojas,ferziger}. 
In a way similar to elastic neutron or  X-ray scattering in liquids  
\cite{lovesey}, the scattering cross section of sound waves  
can be related within the first Born approximation to the modulus of the  
Fourier transform of the vorticity field.   
This result has been experimentally checked for regular laboratory 
flows \cite{baudet,oljaca}, and there is increasing interest in applying 
this method for the study of turbulence \cite{baudet2,pinton,baudet3}. 
In a related development, time-reversal acoustic mirrors have been used to 
probe
vortical flows \cite{fink}.
 
In the present article, we analyze this scattering problem from a 
distribution  
function point of view, picturing the vorticity distribution as an assembly 
of  
many undeformable vortices whose structure is characterized {\it a priori} 
by a given set of particle distribution densities.  
We will consider not only the scattering cross section problem, 
but also the dispersion relation (or effective wave number)  
of an acoustic wave traveling through such a  
medium. We consider a plane wave that propagates in an infinite  
statistically homogeneous flow composed of  
vortices with number density $n$. According to multiple scattering theory  
(see e.g. refs.\cite{frisch,ishimaru,ostbook}), such a medium can be 
described on average by an  
effective  
index of refraction, which generally depends on the wave frequency. 
Moreover, the wavenumber that characterizes the mean wave 
has an imaginary part, 
equal to half the total scattering cross section per unit volume; this 
accounts for the attenuation of the wave amplitude due to the 
loss of coherence during the scattering processes. In order to get as much 
insight as possible while keeping algebraic complications to a minimum we take 
all vortices with the same absolute value of circulation.
 
In a recent publication \cite{pfaI} we derived a general dispersion relation 
for a sound wave propagating in a disordered 
flow of mean velocity zero, and this result was applied to a "gas" of  
statistically independent identical vortices.  
In the next section, we extend the expression of the dispersion relation 
for a population of identical vortices of arbitrary spatial correlations. 
In section III, we specify the study to two dimensional systems  
and underline some formal 
analogies with ionic liquids. 
The results are discussed in section IV, especially in connection with  
two-dimensional decaying turbulence. Technical details 
are given in the Appendix.

\section{"Liquids" of vortices} 
 
The treatment of the interaction between a sound wave and a flow 
presented in ref.\cite{pfaI} follows an usual scheme, already presented  
elsewhere \cite{lundrojas}. The analysis is based on a linearization 
of the Euler and continuity equations for a fluid, and 
relies on several assumptions: 
the velocities associated with the acoustic wave are  
supposed to be much smaller than the typical  
velocities of the base flow, themselves much lower than the speed of sound. 
In this case the acoustic wave is a first order effect over the base flow, that 
can be considered as incompressible to zeroth order.
The frequency $\nu_0$ of the incoming wave is also supposed 
much higher than the frequencies associated with the flow, which is 
considered as frozen, and viscous effects can be neglected. The medium is 
then stationary and averages are made over flow configurations. 
In the following, we denote $c_0$ the speed of sound and $k_0=\nu_0/c_0$  
the wavenumber of the acoustic wave in the flow at rest. 
 
If the flow is composed of $N$ undeformable 
vortices, its velocity field is the sum of the flow fields generated by each  
vortex, 
\begin{equation}\label{uv} 
\vec{u}(\vec{x})=\sum_{a=1}^{N}\vec{v}^{(a)}(\vec{x})\ . 
\end{equation} 
The incident wave is scattered at the velocity inhomogeneities of the medium  
and when averages over disorder configurations are performed, 
the mean acoustic pressure that results from these scattering processes 
can be derived within the framework of multiple scattering  
theory \cite{frisch,ishimaru,ostbook,lax,keller}. The mean wave is 
the sum of coherent scattering paths and  
has the same direction of propagation than the incident wave, noted 
$\hat{k}_0$, 
but is described by an effective wavenumber $k\neq k_0$. 
For a weakly perturbed medium, {\it i.e.} for a flow of low Mach number  
${\cal M}=|\vec{u}|/c_0$ (the vortex density $n$ can be high), $k$ can be  
expanded perturbatively around $k_0$ in multiple scattering series.  
Since the flow has no mean velocity, the term of order ${\cal M}$ is 
identically zero. 
The first non vanishing corrections involve two diagrams of order  
${\cal M}^2$, and the dispersion relations can be written in Fourier space 
as \cite{pfaI}: 
\begin{eqnarray}\label{dispgen} 
k^2&=&k_0^2+k_0^2\ \frac{(\gamma-4)}{c_0^2}\ \frac{1}{V} 
\int\frac{d\vec{q}}{(2\pi)^d}\sum_{a,b}\left\langle 
[\hat{k}_0.\vec{v}^{(a)}(\vec{q})] 
[\hat{k}_0.\vec{v}^{(b)}(-\vec{q})]\right\rangle\\ 
&&+k_0^2\ \frac{4}{c_0^2}\ \frac{1}{V} 
\int\frac{d\vec{q}}{(2\pi)^d}\ \sum_{a,b}\left\langle 
[\hat{k}_0.\vec{v}^{(a)}(\vec{q}-\vec{k}_0)] 
[\hat{k}_0.\vec{v}^{(b)}(\vec{k}_0-\vec{q})]\right\rangle 
(\hat{k}_0.\vec{q})^2{\cal G}_{k_0}^{(0)}(\vec{q})\ ,\nonumber 
\end{eqnarray} 
where $\gamma=c_p/c_v$ is 
the ratio of specific heats at constant pressure and constant volume, 
$V$ the volume occupied by the flow, $d$ the space dimension and 
\begin{equation} 
{\cal G}_{k_0}^{(0)}(\vec{q})=\lim_{\eta\rightarrow0^+}\ 
\frac{1}{q^2-(k_0+i\eta)^2} 
\end{equation} 
is the Fourier transform of the free-space Green's function. 
The second term of the right-hand side of Eq.(\ref{dispgen}) is real and 
proportional to the kinetic energy of the flow per unit volume. The real 
part 
of the last term of the right-hand side is also of order of the kinetic 
energy  
per unit volume, but introduces dispersion. This last term has 
an imaginary part, and using the residue theorem, one deduces that 
${\cal I}m(k)\equiv\Lambda^{-1}$ is given by 
\begin{equation}\label{siggen} 
\Lambda^{-1}=\frac{\sigma^*_T}{2}=\frac{\pi k_0^{d+1}}{c_0^2}\ \frac{1}{V} 
\int\frac{d\Omega^{(d)}}{(2\pi)^d}\ \sum_{a,b}\left\langle 
[\hat{k}_0.\vec{v}^{(a)}(k_0\hat{q}-\vec{k}_0)] 
[\hat{k}_0.\vec{v}^{(b)}(\vec{k}_0-k_0\hat{q})]\right\rangle 
(\hat{k}_0.\hat{q})^2\ , 
\end{equation} 
where $\sigma^*_T$ is the total scattering cross section per unit volume, 
$\hat{q}=\vec{q}/|\vec{q}|$, $d\Omega^{(d)}$ is the solid angle measure, 
and $\Lambda$ is then the attenuation length of the coherent 
wave \cite{sigma}.  It is assumed that the attenuation per wavelength is 
small 
($\Lambda k_0\gg1$), a condition that is fulfilled for a wide range of  
wavelengths if the Mach number is small.  
(See ref.\cite{pfaII} 
for a fuller discussion of the validity of 
Eqs.(\ref{dispgen})-(\ref{siggen})). 
 
If the vortices are all identical, the Fourier transform  
$\vec{v}^{(a)}(\vec{q})$ of the flow field generated by any vortex $(a)$ 
has the form: 
\begin{equation}\label{vortid} 
\vec{v}^{(a)}(\vec{q})=\int d\vec{x}\ \vec{v}(\vec{x}-\vec{x}_a,\hat{a})\ 
e^{-i\vec{q}.\vec{x}}\ , 
\end{equation} 
where $\vec{x}_a$ is the position of the vortex
and $\hat{a}$ an angle describing its global orientation;
$\vec{v}$ is a fixed velocity function that depends on the  
vortex shape. 
The terms in Eq.(\ref{dispgen}) are averaged over all the 
positions and orientations $\{\vec{x}_a,\hat{a}\}$. 
 
In ref.\cite{pfaI}, we have supposed that the cross-terms in the sums of  
Eqs.(\ref{dispgen})-(\ref{siggen}) 
vanish. This situation is encountered either when the flow fields of the  
vortices $(a)$ and $(b)$ do not overlap 
($\vec{v}^{(a)}.\vec{v}^{(b)}\simeq0$), 
{\it i.e.} if the number density $n=N/V$ is small, or if the density is high 
but the vortices are spatially uncorrelated ($\langle 
\vec{v}^{(a)}.\vec{v}^{(b)}\rangle\simeq0$, or mean-field-like assumption). 
 
If correlations between vortices are present, the off-diagonal 
terms of the Born integrals of Eqs.(\ref{dispgen})-(\ref{siggen})  
do not vanish. 
It is thus convenient to introduce  
a pair correlation function for the rigid vortices, noted $g$.  
With a normalized angle measure, 
this distribution function is such that 
$n^2g(\vec{x}_a,\vec{x}_b,\hat{a},\hat{b})d\vec{x}_ad\vec{x}_bd\hat{a} 
d\hat{b}$ represents the number of pairs per unit 
volume and per unit angle formed by two vortices in configurations  
$\{\vec{x}_a,\hat{a}\}$ and  
$\{\vec{x}_b,\hat{b}\}$, respectively.  
Consequently, 
\begin{eqnarray}\label{md0} 
\left\langle v_i(\vec{x}-\vec{x}_a,\hat{a})v_j(\vec{x}-\vec{x}_b,\hat{b}) 
\right\rangle= 
\frac{1}{V^2}\int d\vec{x}_ad\vec{x}_bd\hat{a}d\hat{b}\  
g(\vec{x}_a,\vec{x}_b,\hat{a},\hat{b}) 
v_i(\vec{x}-\vec{x}_a,\hat{a})v_j(\vec{x}-\vec{x}_b,\hat{b}) . 
\end{eqnarray} 
For homogeneous systems with vanishing correlations at large distances,  
$g(\vec{x}_a,\vec{x}_b,\hat{a},\hat{b})= 
g(\vec{x}_a-\vec{x}_b,\hat{a},\hat{b})$, 
and  $g\rightarrow 1$ when $|\vec{x}_a-\vec{x}_b|\rightarrow\infty$. 
Using the fact that $\int d\vec{x}_a v_i(\vec{x}-\vec{x}_a)=0$,  
expression (\ref{md0}) remains unchanged if $g$ is replaced by the  
total correlation function $h\equiv g-1$, which has the advantage 
of a well-defined Fourier  
transform. Using the identity (\ref{md0}), the dispersion relation 
(\ref{dispgen}) can be recast as: 
\begin{eqnarray}\label{dispcorr} 
k^2&=&k_0^2+\frac{(\gamma-4)k_0^2}{c_0^2} 
\int\frac{d\vec{q}}{(2\pi)^d}d\hat{a}d\hat{b}\  
[\hat{k}_0.\vec{v}(\vec{q},\hat{a})][\hat{k}_0.\vec{v}(-\vec{q},\hat{b})] 
\left\{n\delta(\hat{a}-\hat{b})+n^2h(\vec{q},\hat{a},\hat{b})\right\}\\ 
&&+\frac{4k_0^2}{c_0^2} \int\frac{d\vec{q}}{(2\pi)^d}d\hat{a}d\hat{b}\ 
[\hat{k}_0.\vec{v}(\vec{q}-\vec{k}_0,\hat{a})] 
[\hat{k}_0.\vec{v}(\vec{k}_0-\vec{q},\hat{b})] 
(\hat{k}_0.\vec{q})^2{\cal G}_{k_0}^{(0)}(\vec{q})\nonumber\\ 
&&\left\{n\delta(\hat{a}-\hat{b})+n^2h(\vec{q}-\vec{k}_0, 
\hat{a},\hat{b})\right\}\ ,\nonumber 
\end{eqnarray} 
where $\vec{v}(\vec{q},\hat{a})$ is the Fourier transform of the flow field 
of the vortex located at the origin, 
$\vec{v}(\vec{q},\hat{a})=\int d\vec{x}\ \vec{v}(\vec{x},\hat{a}) 
\exp(-i\vec{q}.\vec{x})$. In the same way, we get from Eq.(\ref{siggen}): 
\begin{eqnarray}\label{sigcorr} 
\sigma^*_T&=&\frac{2\pi k_0^{d+1}}{c_0^2} 
\int\frac{d\Omega^{(d)}}{(2\pi)^d}d\hat{a}d\hat{b}\  
[\hat{k}_0.\vec{v}(k_0\hat{q}-\vec{k}_0,\hat{a})] 
[\hat{k}_0.\vec{v}(\vec{k}_0-k_0\hat{q},\hat{b})] 
(\hat{k}_0.\hat{q})^2\\ 
&&\left\{n\delta(\hat{a}-\hat{b})+n^2h(k_0\hat{q}-\vec{k}_0, 
\hat{a},\hat{b})\right\}.\nonumber 
\end{eqnarray} 
(\ref{sigcorr})  
When the vortices are uncorrelated, $h=0$, we recover the results of 
\cite{pfaI,pfaII}.

\section{Two dimensional case} 
 
Two-dimensional turbulence has been the subject of many studies because 
of its possible applications in meteorology and oceanography, but also 
because it is the most accessible dimension for computational and 
theoretical 
approaches. 
In the present context, the two-dimensional study of the effects  
of vortex correlations on sound propagation is clearly simpler  
because of the reduced number of degrees of freedom involved.  
However, we hope that it 
can provide at least some first answers qualitatively valid in any  
dimension, besides the fact that the structure of turbulence itself deeply  
changes with the space dimensionality. 
In two dimensions 
the vorticity $\vec{\omega}$ points in the perpendicular direction, along 
the  
$z$-axis. 
We consider a system composed of axisymmetric vortices, where 
each vortex produces an azimuthal 
velocity field (see ref.\cite{pfaII} for an example) around a vortex core 
of circulation modulus 
$\Gamma$. Suppose that each vortex has a probability $x_+$ to have a  
circulation $+\Gamma$, and a probability $x_-$ to have circulation $-\Gamma$  
($x_+ + x_-=1$). 
The two-particle distributions are 
described by introducing three pair correlation functions, 
$h_{++}$, $h_{--}$ and $h_{+-}$, where the sign of the subscript refers 
to the respective signs of the vortex circulations. 
At this point, it is convenient to introduce distribution functions used 
in the formalism of ionic liquids \cite{hansen}.  
Let us consider the local vortex density, namely a "charge" density,  
defined as 
\begin{equation} 
\rho^Z(\vec{r})=\sum_{i}z_i \rho_i(\vec{r})\ , 
\end{equation} 
where the sum runs over the two species of number density 
$\rho_i(\vec{r})$ and charge $z_i$ (here, the orientation $\pm 1$).  
The vortex, or charge, structure factor $S_{ZZ}$ associated with the density  
$\rho^Z(\vec{r})$, defined by 
\begin{equation} 
S_{ZZ}(\vec{q})= 
\frac{\langle \rho^Z(\vec{q})\rho^Z(-\vec{q})\rangle}{N}\ , 
\end{equation} 
can be rewritten as 
\begin{equation}\label{szzh} 
S_{ZZ}(\vec{q})=1+n\left[ 
x_+^2 h_{++}(\vec{q})+x_-^2 h_{--}(\vec{q})-2x_+ x_- h_{+-}(\vec{q})\right] 
. 
\end{equation} 
From Eq.(\ref{dispcorr}), we can deduce the expression for  
the index of refraction defined as  
${\cal N}=c/c_0={\cal R}e(k_0/k)$, where $c$ denotes the new phase velocity. 
Noting that  
$\vec{v}(\vec{x},-\hat{z})=-\vec{v}(\vec{x},\hat{z})$, and assuming that the  
flow is isotropic, one gets 
\begin{eqnarray}\label{dispcorr2d} 
\delta{\cal N}&=&1-\frac{n(\gamma-4)}{8\pi c_0^2}\int_0^{\infty} dq\ q\   
v^2(q)S_{ZZ}(q)\\ 
&&-\frac{2n}{c_0^2}\ {\cal R}e\ \int\frac{d\vec{q}}{(2\pi)^2} 
\left|\hat{k}_0.\vec{v}(|\vec{q}-\vec{k}_0|)\right|^2  
S_{ZZ}(|\vec{q}-\vec{k_0}|) (\hat{k}_0.\vec{q})^2{\cal G}^{(0)}_{k_0}(q). 
\nonumber 
\end{eqnarray} 
As a consequence of the structure of Eq.(\ref{dispcorr2d}), 
the asymptotic results obtained in ref.\cite{pfaII} for the index of  
refraction in the limits of short and long wavelengths,   
remain valid here, provided that one replaces any quadratic factor $v^2$ by  
$v^2 S_{ZZ}$.  
 
The total scattering cross-section, in turn, is given by: 
\begin{equation}\label{sigcorr2d} 
\sigma_T^*=\frac{nk_0^3}{2\pi c_0^2}\int d\Omega^{(2)} 
\left|\hat{k}_0.\vec{v}(|k_0\hat{q}-\vec{k}_0|)\right|^2  
S_{ZZ}(|k_0\hat{q}-\vec{k_0}|) (\hat{k}_0.\hat{q})^2. 
\end{equation} 
Using the property $\vec{q}.\vec{v}(\vec{q})=0$ for incompressible 
fluids (a correct assumption for low Mach number base flows), as illustrated 
by the geometrical construction of Figure 1, the above  
expression can be rewritten 
\begin{equation}\label{sigcorr2dbis} 
\sigma_T^*=\frac{nk_0^3}{2\pi c_0^2}\int_{-\pi}^{\pi}d\theta\  
\frac{\sin^2\theta\cos^2\theta}{2(1-\cos\theta)}\  
v^2\left(k_0\sqrt{2-2\cos\theta}\right)  
S_{ZZ}\left(k_0\sqrt{2-2\cos\theta}\right) . 
\end{equation} 
With the same techniques presented in ref.\cite{pfaII}, it is easy to show  
that $\sigma_T^*\sim k_0^2$ when $k_0\rightarrow\infty$. 
If $S_{ZZ}(q)$ behaves like $q^\beta$ at small $q$, one gets, 
from Eq.(\ref{sigcorr2dbis}), 
\begin{equation}\label{sig0} 
\sigma_T^*\sim k_0^{5+\beta} ,\quad k_0\rightarrow 0 , 
\end{equation} 
where we have used the property $v(q\rightarrow0)\sim q$ for an axisymmetric  
bounded flow.

\section{Discussion} 
 
Relations (\ref{dispcorr2d}) and (\ref{sigcorr2d}) show that the coherent  
propagation of an acoustic wave through a population of identical vortices  
closely depends on their spatial structure.  
In two dimensions, the index of refraction and the attenuation length 
of the wave involve a circulation  
(or "charge", by reference to ionic liquids) structure factor. Hence, 
if the vortices have a simple shape and generate a flow field $v$ with well  
known characteristics, information on this distribution function 
can be deduced from the study of the acoustic properties mentioned above. 
Tractable analytical expressions  
have only been obtained in two dimensions. However, we think that they 
provide a qualitative understanding of the probe of vortex correlations 
by ultrasound techniques, even in higher space dimensions.  
 
For its similarity with two-dimensional homogeneous turbulence\cite{2dtur}, let 
us 
consider further the case of a neutral flow, where $x_+=x_-=1/2$. 
There are mainly two ways of considering the vortex structure 
of decaying two-dimensional turbulence, since there are two  
very different length scales in the problem. Depending on the scale of  
interest, one can either picture the flow as composed of many  
small, nearly point-like vortices that may form large structures, or   
consider the flow as formed by these few large coherent vortices only. 
One can assume that the size of the former 
vortices is of order of a dissipation scale, while the size of the latter  
is of order of the integral scale, at which energy is initially injected. 
 
An ideal gas of vortices is characterized by $S_{ZZ}(q)=1$; 
however, Eq.(\ref{szzh}) shows that it is equivalent,  
with respect to the dispersion relation, to identical  
ordering between identical and opposite vortices ($h_{++}+h_{--}=2 
h_{+-}$). We logically check that a tendency to observe vortex  
pairs of identical circulation ($h_{++}+h_{--}>2h_{+-}$) would result in an 
increase of the kinetic energy of the flow, and hence the phase change. 
The limit of $S_{ZZ}(q)$ at small wavenumbers is related to that of 
the total scattering cross section through relation (\ref{sig0}) if the flow 
produced by one vortex is bounded in space. 
$S_{ZZ}(0)$ indeed represents the local fluctuations of the number of 
charges,  
and is of particular interest. A local electroneutrality  
assumption ($S_{ZZ}(0)=0$) means that  
the "charge" (or circulation) of a given vortex is  
exactly canceled by the total "charge" of the vortices that 
surround it: at length scales of the order of a few vortex radii,  
the fluid has no net rotation.  
 
Let us consider the flow at small scales, 
{\it i.e.} made of nearly point-like vortices. 
Although not much studied 
in the literature, vortex pair correlations have motivated theoretical 
works. 
An exact analytical expression for the vorticity structure factor 
has been derived for a class of two-dimensional stationary solutions  
of the Navier-Stokes equation\cite{cook}. It is characterized by a 
Debye-H${\rm \ddot{u}}$ckel-like pair distribution, 
$S_{ZZ}(q)\propto q^2/(q^2+k_s^2)$. 
Turbulence at high Reynolds number clearly exhibits quite distinct  
statistical properties: computer simulations rather show that 
vortices with same circulation sign have the tendency to organize 
in domains \cite{montgomery}. 
Such systems, where electroneutrality is not locally 
observed, are theoretically better described by 
other approaches, like the microcanonical formulations of the statistics  
of two dimensional vortices in a bounded domain 
\cite{onsager,novikov,pointin,rosom,miller}. 
These theories predict 
that the one-point probability 
distribution of vortices is spatially nonuniform. 
To obtain this result, however,  
two-point correlation functions are usually neglected at first order 
($S_{ZZ}=1$).  
 
Two-dimensional homogeneous isotropic turbulence 
can be conveniently described in the inertial range by uniform mixture of  
long range correlated vortices. 
Notice that the two-dimensional energy spectrum, 
$E(q)=q\langle\vec{u}(\vec{q}). 
\vec{u}(-\vec{q})\rangle/(4\pi V)$, can be reexpressed with the charge  
structure factor as $E(q)=n/(4\pi)q\ v^2(q)S_{ZZ}(q)$. 
The interpretation of the classical two-dimensional 
turbulence spectral laws $E(k)\sim k^{-\mu}$,  
($\mu=3$ \cite{soap,batchelor,kraichnan2}, $\mu\simeq4$ \cite{benzi,deem})  
in terms of quasi point-like vortices ($v(q)\sim q^{-1}$) 
leads to $S_{ZZ}\sim q^{-(\mu-1)}$. This is supposed to be the 
behavior of $S_{ZZ}$ in the inertial range.  
When the structure factor is steep enough (say, if $\mu$ is significantly  
larger than 1), one expects the vorticity correlations to be long range and 
positive: the spatial structure differs qualitatively 
from that of the ideal gas given by a white spectrum.  
The common spectral laws are compatible with 
preferential ordering between same sign vortices. 
 
At large scales, the flow is made of large coherent vortical structures. 
The pair correlations between these vortices  
are closely related to the properties of the 
energy spectrum in the low wavenumber limit, outside of the inertial range. 
For instance, if the spectrum in this limit is such that $E(q)\sim 
q^{\mu'}$,  
with $\mu'>0$ \cite{lesieur}, the electroneutrality is local beyond  
the size of the coherent vortices. In that 
case, one expects that 
the structure of the vortex system should be similar to the 
short range structure in liquids.

\acknowledgments{ The authors are grateful to G. Tarjus for fruitful  
discussions. 
  This work was supported by Fondecyt Grants  
  3970013, 1960892, and a  
  C\'atedra Presidencial en Ciencias. 
} 
 
 
\appendix 
 
\section{} 
 
For identical vortices and with the help of relation (\ref{vortid}),  
the diagonal terms $(a)=(b)$ of the sum 
\begin{equation} 
\frac{1}{V}[\hat{k}_0.\vec{u}(\vec{q})][\hat{k}_0.\vec{u}(-\vec{q})]= 
\frac{1}{V}\sum_{a,b}\left\langle 
[\hat{k}_0.\vec{v}^{(a)}(\vec{q})] 
[\hat{k}_0.\vec{v}^{(b)}(-\vec{q})]\right\rangle 
\end{equation} 
give the contribution 
\begin{equation}\label{td} 
n\int d\hat{a}\  
[\hat{k}_0.\vec{v}(\vec{q},\hat{a})][\hat{k}_0.\vec{v}(-\vec{q},\hat{a})]\ , 
\end{equation} 
where $\vec{v}(\vec{q},\hat{a})=\int d\vec{x}\ \vec{v}(\vec{x},\hat{a}) 
\exp(-i\vec{q}.\vec{x})$.  
The off-diagonal terms $(a)\neq (b)$ take the form 
\begin{equation} 
\frac{N^2}{V}\frac{1}{V^2}\int d\vec{x}d\vec{x}'d\vec{x}_a d\vec{x}_b 
d\hat{a}d\hat{b}\ g(\vec{x}_a-\vec{x}_b,\hat{a},\hat{b}) 
e^{-i\vec{q}.(\vec{x}_a-\vec{x}_b)} 
\hat{k}_0.\vec{v}(\vec{x},\hat{a})e^{-i\vec{q}.\vec{x}} 
\hat{k}_0.\vec{v}(\vec{x}',\hat{b})e^{i\vec{q}.\vec{x}'} 
\end{equation} 
and can be recast as 
\begin{equation}\label{tnd} 
n^2\int d\hat{a}d\hat{b}\  
[\hat{k}_0.\vec{v}(\vec{q},\hat{a})][\hat{k}_0.\vec{v}(-\vec{q},\hat{b})] 
h(\vec{q},\hat{a},\hat{b}). 
\end{equation} 
Summing the terms (\ref{td}) and (\ref{tnd}), the relations (\ref{dispgen}) 
and (\ref{siggen}) are transformed to Eqs.(\ref{dispcorr}) and 
(\ref{sigcorr}).

\begin{figure} 
  \caption{ 
    Axisymmetric vortex in two dimensions. 
  } 
  \label{fig1} 
\end{figure} 
 

\begin{thebibliography}{99}
\bibitem{sigma} Formula (\ref{siggen}) for the total scattering cross  
section $\sigma^*_T$ is the same as its expression obtained  
from the scattered amplitude in the first Born  
approximation. $\hat{q}$ represents the direction of the scattered wave. 
This result has already been obtained, but presented in another  
form, in refs.\cite{kraichnan,lundrojas}.
\end{thebibliography}
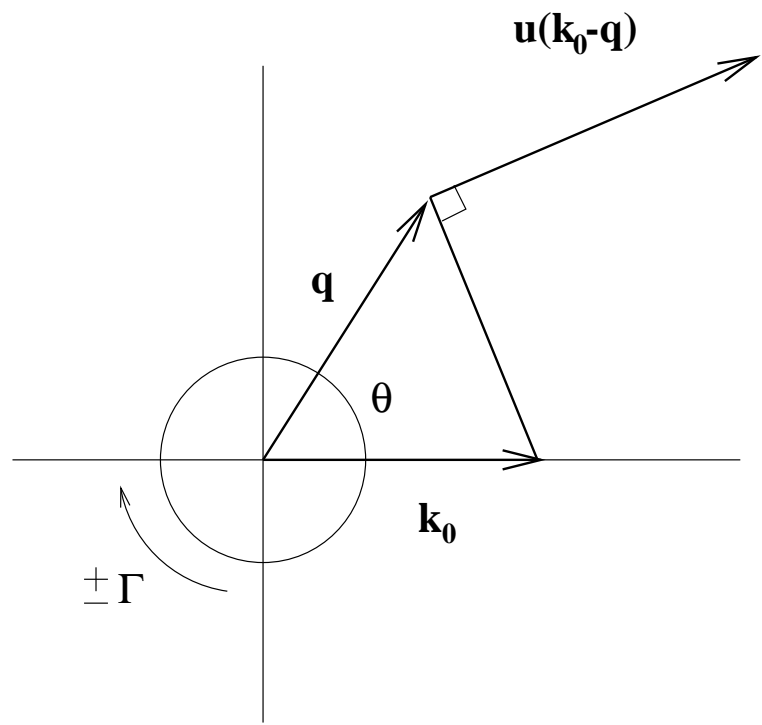
\end{document}